\documentclass[preprint,nohyper]{JHEP3} %
\usepackage{epsfig,multicol}

\newcommand{\newc}{\newcommand}
\newc\eg{{\it {e.g.}}}  \newc\etal{{\it {et al.}}} \newc\ie{{\it i.e.}}
\newc\etc{{\it {etc}}}  

\newcommand\lsim{\mathrel{\rlap{\lower4pt\hbox{\hskip1pt$\sim$}}
    \raise1pt\hbox{$<$}}}
\newcommand\gsim{\mathrel{\rlap{\lower4pt\hbox{\hskip1pt$\sim$}}
    \raise1pt\hbox{$>$}}}

\newc{\mhalf}{m_{1/2}}      \newc{\mzero}{m_0}

\newc{\tanb}{\tan\beta}
\newc{\azero}{A_0}
\newc{\at}{A_t} \newc{\abot}{A_b} \newc{\atau}{A_\tau} 

\newc{\bmu}{B\mu}           \newc{\sgn}{{\rm sgn}}
\newc{\mone}{M_1}           \newc{\mtwo}{M_2}

\newc{\charone}{\chi_1^\pm} \newc{\mcharone}{m_{\chi_1^\pm}}

\newc{\hl}{h}               \newc{\mhl}{m_{\hl}}
\newc{\hh}{H}               \newc{\mhh}{m_{\hh}}
\newc{\ha}{A}               \newc{\mha}{m_{\ha}}
\newc{\hc}{H^{\pm}}         \newc{\mhc}{m_{\hc}}

\newc{\mgut}{M_{\rm GUT}}

\newc{\mplanck}{M_{\rm P}}      \newc{\mpl}{M_{\rm Pl}}
\newc{\msusy}{M_{\rm SUSY}}      \newc{\ms}{M_{\rm S}}

\newc{\jxf}{J({\xf})}
\newc{\jxfexact}{J_{\rm exact}({\xf})}  \newc{\jxfexp}{J_{\rm exp}({\xf})}
\newc{\VEV}[1]{\langle #1 \rangle}

\newc{\xf}{x_f}
\newc\vrel{v_{\rm rel}}

\newc\sell{{\widetilde e}_L}      \newc\msell{m_{\sell}}
\newc\selr{{\widetilde e}_R}      \newc\mselr{m_{\selr}}

\newc\snue{{\widetilde \nu}_e}      \newc\msnue{m_{\snue}}
\newc\snutau{{\widetilde \nu}_\tau}      \newc\msnutau{m_{\snutau}}

\newc\supl{{\widetilde u}_L}      \newc\msupl{m_{\supl}}
\newc\supr{{\widetilde u}_R}      \newc\msupr{m_{\supr}}
\newc\sdl{{\widetilde d}_L}      \newc\msdl{m_{\sdl}}
\newc\sdr{{\widetilde d}_R}      \newc\msdr{m_{\sdr}}

\newcommand\stauone{{\tilde \tau}_1}   \newcommand\mstauone{m_{\stauone}}

\newcommand\gluino{\tilde g}
\newcommand\mgluino{m_{\gluino}}

\newc\hpm{H^\pm} \newc\hp{H^+} \newc\hm{H^-} 
\newc\sfermion{\tilde f}  \newc\msfermion{m_{\sfermion}}  

\newc\second{{\rm sec}} 

\newc\alphas{\alpha_s}

\newc\alphaem{\alpha_{em}}

\newcommand\mz{m_{Z}}

\newcommand\treh{T_{\rm R}}

\newc{\sthw}{\sin\theta_W}              \newc{\cthw}{\cos\theta_W}
\newc{\bino}{\widetilde B}              \newc{\wino}{\widetilde W_3}
\newc{\higgsinob}{{\widetilde H}^0_b}   \newc{\higgsinot}{{\widetilde H}^0_t}

\newc{\abund}{\Omega h^2}
\newc{\abundchi}{\Omega_\chi h^2}
\newc{\abundcdm}{\Omega_{{\rm CDM}} h^2}
\newc{\omegam}{\Omega_{{\rm M}}}       \newc{\abundm}{\Omega_{{\rm M}} h^2}
\newc{\omegab}{\Omega_{{\rm b}}}	\newc{\abundb}{\Omega_{{\rm b}} h^2}
\newc{\omegacdm}{\Omega_{{\rm CDM}}}   \newc{\omegatot}{\Omega_{{\rm TOT}}}

\newc{\rhocrit}{\rho_{crit}}
\newc{\rhochi}{\rho_{\chi}}

\newcommand\fa{f_{a}}

\newcommand\stau{\tilde{\tau}}

\newcommand\neut{\tilde \chi}
\newcommand\qbar{\bar q}
\newc{\cachigamma}{C_{a\neut\gamma}}
\newc{\caww}{C_{aWW}}                   
\newc{\cayy}{C_{aYY}}

\newc{\nl}{\cos \theta_{\tilde t}}
\newc{\nr}{\sin \theta_{\tilde t}}
\newcommand\tev{\,\mbox{TeV}}
\newcommand\gev{\,\mbox{GeV}}
\newcommand\mev{\,\mbox{MeV}}
\newcommand\kev{\,\mbox{keV}}
\newcommand\ev{\,\mbox{eV}}

\newc\gbar{{\overline{g}}}

\newc{\ra}{\rightarrow}

\newc{\beq}{\begin{equation}}
\newc{\eeq}{\end{equation}}
\newc{\bea}{\begin{eqnarray}}
\newc{\eea}{\end{eqnarray}}

\renewcommand\({\left(}
\renewcommand\){\right)}
\renewcommand\[{\left[}
\renewcommand\]{\right]}

\newc{\nspin}{n_{\rm spin}}
\newc{\nflavor}{n_{\rm F}}
\newc{\ngamma}{n_\gamma}
\newc{\ychi}{Y_{\chi}}                  \newc{\yeqchi}{Y^{\rm EQ}_{\chi}}

\newcommand\axino{\tilde{a}}        
\newcommand\maxino{m_{\axino}}
\newcommand\abunda{\Omega_{\axino}h^2}

\newc{\naxino}{n_{\axino}}
\newc{\yaxino}{Y_{\axino}}
\newc{\yeqaxino}{Y^{\rm EQ}_{\axino}}
\newc{\ythaxino}{Y^{\rm TP}_{\axino}}
\newc{\ynthaxino}{Y^{\rm NTP}_{\axino}}
\newc{\yascat}{Y^{\rm scat}_{i,j}}      \newc{\yadec}{Y^{\rm dec}_{i}}
\newc{\gstar}{g_\ast}           \newc{\gsstar}{g_{s\ast}}

\def\lag             {{\cal L}}

       \def\pslash{\not{\hbox{\kern-2.3pt $p$}}}
       \def\kslash{\not{\hbox{\kern-2.3pt $k$}}}
       \def\qslash{\not{\hbox{\kern-2.3pt $q$}}}
       \def\ddslash{\not{\hbox{\kern-2.3pt $d$}}}
       \def\prtslash{\not{\hbox{\kern-2.3pt $\partial$}}}

\title{Axino Dark Matter and the CMSSM}

\author{Laura Covi\\
TH Division, Dept. of Physics, CERN, CH-1211 Geneva 23, Switzerland
}
\author{Leszek Roszkowski, Roberto Ruiz de Austri\\
Department of Physics and Astronomy,
University of Sheffield, Sheffield, S3 7RH, England
}
\author{Michael Small\\
KPMG, Preston, PR2 2YF, England
}

\abstract{
If the axino is the lightest superpartner and satisfies cosmological
bounds, including a preferred range of the relic abundance of cold
dark matter, then the usual stringent constraints on the parameter
space of the CMSSM become greatly relaxed. The lightest superpartner of the
usual CMSSM spectrum will appear to be stable in collider experiments but
will not necessarily obey relic abundance constraints. It may be
either neutral (lightest neutralino) or charged (typically a
stau). With the axino as cold dark matter, large regions of the CMSSM,
often corresponding to heavy superpartners, become allowed, depending
on the axino mass and the reheating temperature.
}

\preprint{CERN--PH--TH/2004--023\\
hep-ph/0402240
}
\keywords{Supersymmetric Effective Theories, Cosmology of
Theories beyond the SM, Dark Matter, Supersymmetric Standard Model}
 
\begin{document}

\section{Introduction}\label{sect:intro}

The lightest supersymmetric particle (LSP) that arises in models
featuring supersymmetry (SUSY) at low energy with the conservation of
$R$--parity can be a cold dark matter (DM) candidate~\cite{MSSM-dm}.
The requirement that the LSP is neutral and produced in the required
abundance in the early Universe sets strong constraints on the minimal
supersymmetric standard model (MSSM)~\cite{MSSM-dm} or its constrained
version (CMSSM)~\cite{CMSSM-dm}.  
In the CMSSM, the LSP
is often the lightest neutralino with a large bino
component~\cite{na92,rr93,kkrw94}. 
In particular, the cases of a neutralino nearly
degenerate with the lighter stau~\cite{efo98} or stop~\cite{eos03}
are often of cosmological importance, but parameter choices 
for which such charged particles themselves are the LSP are thought 
to be excluded on astrophysical
grounds.  In fact, very strong bounds on the presence of electrically
charged and colored relics have been obtained using the unsuccessful
searches for exotic nuclei: strong bounds on electrically charged relics
have been obtained up to masses of the order $10^8$ GeV,
while masses greater than a few $\tev$ for colored thermal relics remain 
consistent with observation~\cite{ch-lsp}.\footnote{
Note that in any case for 
stable charged particles not excluded by searches the overclosure 
bound applies, thus strongly limiting the possibility of such heavy
thermal relics, except if some mechanism for suppressing its  
number density is at work. 
}

However, if the LSP does not belong to the usual MSSM spectrum, other
possibilities arise. One such attractive scenario involves
supersymmetric models implementing the Peccei--Quinn mechanism for
solving the strong CP problem~\cite{pq}.  In this class of models the axino,
the fermionic superpartner of the axion, can naturally be the LSP and
constitute the dominant component of DM in the form of
warm~\cite{kmn,bgm,rtw,ckkr,ay00} or cold dark matter
(CDM)~\cite{ckr,ckkr,crs1}. (See also~\cite{ckl00,kk02}.) 
This is because, unlike for the neutralino, the mass of the axino is
not directly determined by the soft SUSY--breaking terms and can be
much smaller~\cite{axinomass}.

Axinos can be efficiently produced in the early
Universe through several possible processes. A class of {\em thermal
production} (TP) processes involves scatterings and decays of particles in the
primordial plasma. Alternatively, in {\em non--thermal production}
(NTP) the next--to--lightest supersymmetric particle (NLSP) decays
after first freezing out from the plasma. In addition, one can think
of other possible production mechanisms, \eg\ from inflaton decay, but
they are much more model dependent and not necessarily as
efficient. The above processes are supposed to re--generate the
axinos, after their primordial population has been diluted as a result 
of inflation
and subsequent reheating with $\treh\lsim\fa$ (with
$\fa\sim10^{11}\gev$ being the Peccei--Quinn scale) in order to avoid
overclosure. Otherwise, the axinos would have to be very light,
$\maxino<0.2\kev$~\cite{rtw}, with an update in~\cite{ckkr}, thus
constituting warm DM.

In~\cite{ckr,ckkr,crs1} we conducted a detailed investigation of the
axino LSP as CDM in the KSVZ--type (hadronic) axion models~\cite{ksvz}
coupled with the MSSM. As summarized in Figs.~8 and~9 of~\cite{crs1},
we showed that, for $\maxino\gsim 100\kev$ and $\treh\lsim 5\times
10^{6}\gev$, the axino can constitute {\em cold} DM, while at a lower
mass range and larger $\treh$ it could be a warm or even hot DM relic.
At $\treh\gsim 10^{4}\gev$ NTP is typically subdominant while TP
processes allows for a narrow region of $\maxino$ to satisfy the relic
abundance constraint $\abunda\sim1$. On the other hand, in the region
of $\maxino\gsim 10\mev$ and $\treh\lsim 10^{4}\gev$, the NTP
mechanism of neutralino (or other NLSP) decay after freeze-out often
plays a dominant role in producing enough axinos, while at larger
$\treh$ TP does this too efficiently.
Contrary to the case of a gravitino, the other popular alternative
SUSY candidate for the LSP and CDM, the lifetime of the NLSP is much
shorter than $10^4 \sec$ and therefore avoids the very strong bound
coming from photo--destruction of light elements after
nucleosynthesis. For this reason, only the much weaker constraints
from hadronic showers play a role for the axino LSP scenario.
However, if the squarks are much lighter than the gluino, their decays
in TP processes can reduce an upper bound on $\treh$
further~\cite{crs1}.

In this paper, we turn our attention to investigating the implications
on the allowed spectra of the CMSSM of assuming the axino as the LSP
and CDM, with the right amount of the cosmological relic abundance and
with other constraints, especially from nucleosynthesis,
satisfied. This is of interest because the standard paradigm is highly
constrained, especially at not too large values of $\tanb$ on which we
will concentrate here. While this makes the CMSSM highly predictive, a
question arises as to whether viable alternative scenarios exist which
would lead to very different predictions and to possibly relaxing the
stringent cosmological bounds on the CMSSM without abandoning a
supersymmetric explanation for CDM.

As we will see, very different regions of the CMSSM parameter space
will often become allowed.  In particular, it will be possible to have
either an electrically neutral NLSP (the lightest neutralino) or a
charged one (the lighter stau). The NLSP will appear in a collider
detector as a stable lightest ordinary supersymmetric partner
(LOSP). Since axino couplings are suppressed by the Peccei--Quinn
scale, the NLSP lifetime is long compared to the timescales of
relevance for accelerator searches. Therefore, the usual mass bounds
arising from the decay of the sparticle into the lightest neutralino
are not applicable. Depending of the case of a chargino or slepton
L(O)SP, the bounds become a bit relaxed or more stringent.
For the case of a ``stable" charged L(O)SP, a general lower bound 
of about $99.4 \gev$ has been obtained at
LEP~\cite{lep-ch-lsp}.  This may, once superpartners are discovered at
the LHC, allow one to distinguish the case of the axino from the
usually assumed case of the lightest neutralino as the true LSP.

In this context it might be worth commenting on the gravitino as the
LSP and CDM. In a recent study~\cite{eoss04-gravitino} it has been concluded
that one can find limited regions of the CMSSM where the gravitino LSP
would be allowed but that, in these regions, NLSP decays typically 
provide too little relic abundance for the gravitino to fall into the
favored range, if one does not want to spoil nucleosynthesis
predictions.  However, thermal production processes involving gluino
decay and scattering may rectify this \cite{thermal-grav}.

In the following, we will first summarize the effective
axino--neutralino couplings and compute effective axino--stau
couplings and corresponding NLSP lifetimes in the MSSM coupled with
the KSVZ axion models. We will then discuss bounds on the NLSP number
density from primordial nucleosynthesis constraints and summarize the
calculation of axino relic abundance. We will next concentrate on the
CMSSM and will first illustrate the cosmologically allowed parameter
space in the CMSSM in the standard scenario with the neutralino
LSP. Next we will re--compute the relic abundance in the CMSSM with
the axino as the LSP and will investigate the resulting changes in the
cosmological bounds on the ($\mhalf,\mzero$) plane for several values
of the axino mass and $\treh$.

\section{The Axion Multiplet Interactions}

In the Peccei--Quinn (PQ) solution to the strong CP problem, a complex
scalar field is used to break the global $U(1)_{PQ}$ at a high scale
$\fa\sim 10^{11}\gev$.
Its Goldstone boson component, the axion, plays the role of a
dynamical field $\theta_{QCD}$ and relaxes at the origin after the QCD
phase transition~\cite{axion,axionreviews:cite}. After chiral symmetry
breaking, the axion acquires a tiny mass from instanton
effects~\cite{qcdanomaly:cite}.  As before, in this work we
concentrate on the KSVZ--type axion type of interactions \cite{ksvz}.
In the DFSZ case \cite{dfsz} in general the mixing between the
axino and the neutralinos can become substantial and
enhance the couplings discussed here.
 
In the context of supersymmetry the axion field becomes a chiral
multiplet~\cite{susyaxion} which contains not only the pseudoscalar axion
and the scalar saxion with $R$--parity $+1$, but also one $R=-1$ state,
their fermionic superpartner, the axino $\axino$.  The interaction of the axino
with gluons and gluinos proceeds via diagrams that are the
supersymmetric analogues of the fermionic triangle anomaly diagrams
for the heavy states $Q$.  After integrating out the heavy KSVZ quarks
$Q$ and squarks $\widetilde{Q}$, where $m_{Q,\widetilde Q}\sim\fa$, there arises
an effective dimension--5 interaction term
\beq
\label{eq:agg}
\lag_{\axino g \gluino}= \frac{\alpha_s}{8 \pi (\fa/N)}
\bar{\axino} \gamma_5 \sigma^{\mu\nu} \gluino^b G^b_{\mu\nu}
\eeq
where $b=1,\ldots,8$, $\gluino$ is the gluino and $G$ is the strength of
the gluon field, while 
$N$ is the number of flavors of quarks with the Peccei--Quinn
charge, and is equal to $1$ ($6$) in KSVZ (DFSZ) models. For the remainder of
this paper, we understand $\fa$ to mean $\fa/N$.

The coupling~(\ref{eq:agg}) and the corresponding axion coupling
can also be written in a supersymmetric form as a Wess--Zumino term 
in the MSSM superpotential
\beq
W_{WZ} = {\alpha_s \over 2 \sqrt{2} \pi \fa} {\cal A} \;
Tr \left[ W_\alpha W^\alpha \right],
\eeq
where ${\cal A}$ is the axion chiral multiplet, $W_\alpha $ the vector
multiplet containing the gluon, and the trace sums over color indices.
An interesting feature is that the coupling above
is only determined by the QCD anomaly of the heavy states
and is not subject to renormalization~\cite{anom-ren} 
or model dependence.

In an analogous way, Wess--Zumino terms can arise also for 
other gauge interactions,
depending on the charge of the heavy quark multiplet. At high energies,
where all leptons can be considered massless, such interactions can be 
rotated into the $U(1)_Y$ direction~\cite{ckkr} and we are left to
consider only the term
\beq
W_{WZ} = {\alpha_Y \cayy \over 4 \sqrt{2} \pi \fa} {\cal A} \;
 B_\alpha B^\alpha ,
\eeq
where  $B_\alpha $ is the hypercharge vector multiplet and
$ \cayy $ is a model--dependent factor~\cite{coupling}, which vanishes 
if the heavy KSVZ quarks are electrically neutral. If 
they have electric charge $e_Q=-1/3,+2/3$, then $C_{aYY}=2/3,8/3$.
This superpotential term leads to the addition of the following 
effective dimension--5 interaction term to the low--energy Lagrangian
\beq
\label{eq:abb}
\lag_{\axino B \bino}= \frac{\alpha_Y C_{aYY}}{8 \pi \fa}
\bar{\axino} \gamma_5 \sigma^{\mu\nu} \bino B_{\mu\nu}
\eeq
Cosmological implications of the effective axino--gauge boson--gaugino 
operators~(\ref{eq:agg}) and ~(\ref{eq:abb}) have been extensively studied 
in~\cite{ckkr}. 

However, in addition to the above interactions, there is also an
effective dimension--4 coupling of the axino to fermions and
sfermions~\cite{crs1} 
\beq
\lag_{\axino \psi \widetilde \psi}= \Sigma_j\,\,
g_{{\rm eff},\, j}^{L/R}\; \widetilde \psi^{L/R}_j\, \bar \psi_j P_{R/L} \gamma_5 \axino,
\label{eq:sffa}
\eeq
where $\psi_j $ and $\widetilde \psi_j $ can be any of the SM fermions and their
superpartners.
This effective coupling, which arises at two--loop level in KSVZ
models (and therefore at a one--loop level in the effective theory
valid much below $\fa$), is the dominant channel for inducing the decay of
charged NLSPs to axinos.

The effective vertex in~(\ref{eq:sffa}) and its effect on the axino
abundance was computed in the case of light quarks in~\cite{crs1} where
it was found that the dominant contribution was due to 
the logarithmically divergent part of the gluon--gluino--quark loop and was 
proportional to the gluino mass  $\mgluino$,
\beq
g_{{\rm eff},\,  q}^{L/R}
\simeq
\mp {\alpha_s^2 \over \sqrt{2} \pi^2} 
{\mgluino\over \fa} \log\({\fa\over \mgluino}\).
\label{eq:geffapprox-q}
\eeq 
Here the quark mass has been neglected.  
This result was obtained in the low energy
effective theory given by the MSSM plus the interaction
in~(\ref{eq:agg}), \ie\ from the one--loop diagrams as in
Fig.~\ref{fig:sqqa}, where the Peccei--Quinn scale $\fa $ was used as
a cut--off in the loop momentum integral. This procedure is justified
by the fact that the effective theory is valid only below the scale
and in practice amounts to absorbing our ignorance about the high
energy theory into the parameter $\fa$.  In this way a
model--independent computation of the effective coupling was
performed. 
(Integrating the Renormalization Group
Equation for the effective coupling above between $\fa $ and 
$\mgluino $ with vanishing boundary condition, gives very similar
results.)
Adopting the same strategy as in~\cite{crs1}, here we 
extend the computation of the effective vertex to the case of the
tau--stau--axino, including the effects of the tau mass.

Note that the procedure above gives the dominant coupling of the axino 
to the sfermions as long as the mixing of the axino with the other 
neutralinos is negligible. In the DFSZ case, this is not the case
and therefore the computation of the production of axinos in the
thermal bath is more involved. For what concerns NTP instead,
the picture is similar, apart from shorter NLSP lifetimes \cite{martinaxino}.

\FIGURE[t]{%
  \epsfig{file=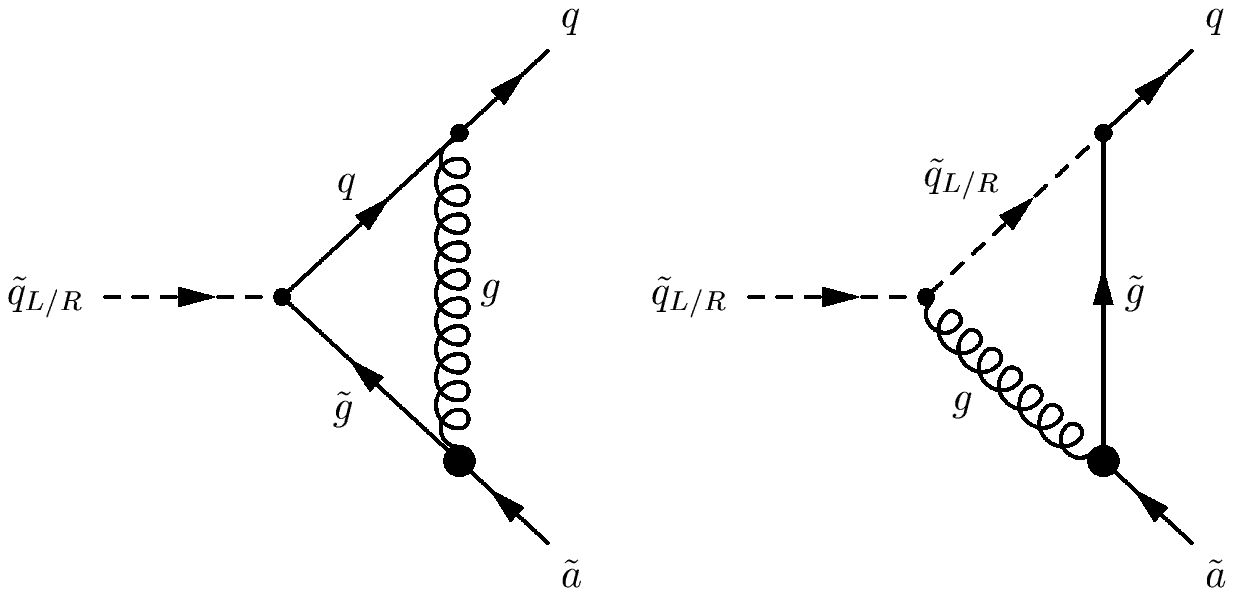, width=5in}
  \caption{The Feynman diagrams contributing to the squark--quark--axino 
            interaction. The thick dot denotes the effective 
             gluino--gluon--axino vertex.}
  \label{fig:sqqa}
}

\section{The Lighter Stau as the NLSP}

In order to study the charged NLSP scenario in the CMSSM, first we evaluate
the effective couplings between the lighter stau and the axino.

\subsection{An Effective Stau--Tau--Axino Coupling}

As shown in Fig.~\ref{fig:loop6}, the stau couples to an axino and a tau via 
triangle diagrams analogous to those giving rise to the squark--quark--axino 
coupling. The effective axino--gauge boson--neutralino vertex in this 
diagram, represented by a thick dot, is given by~(\ref{eq:abb}).
\FIGURE[t]{%
    \epsfig{file=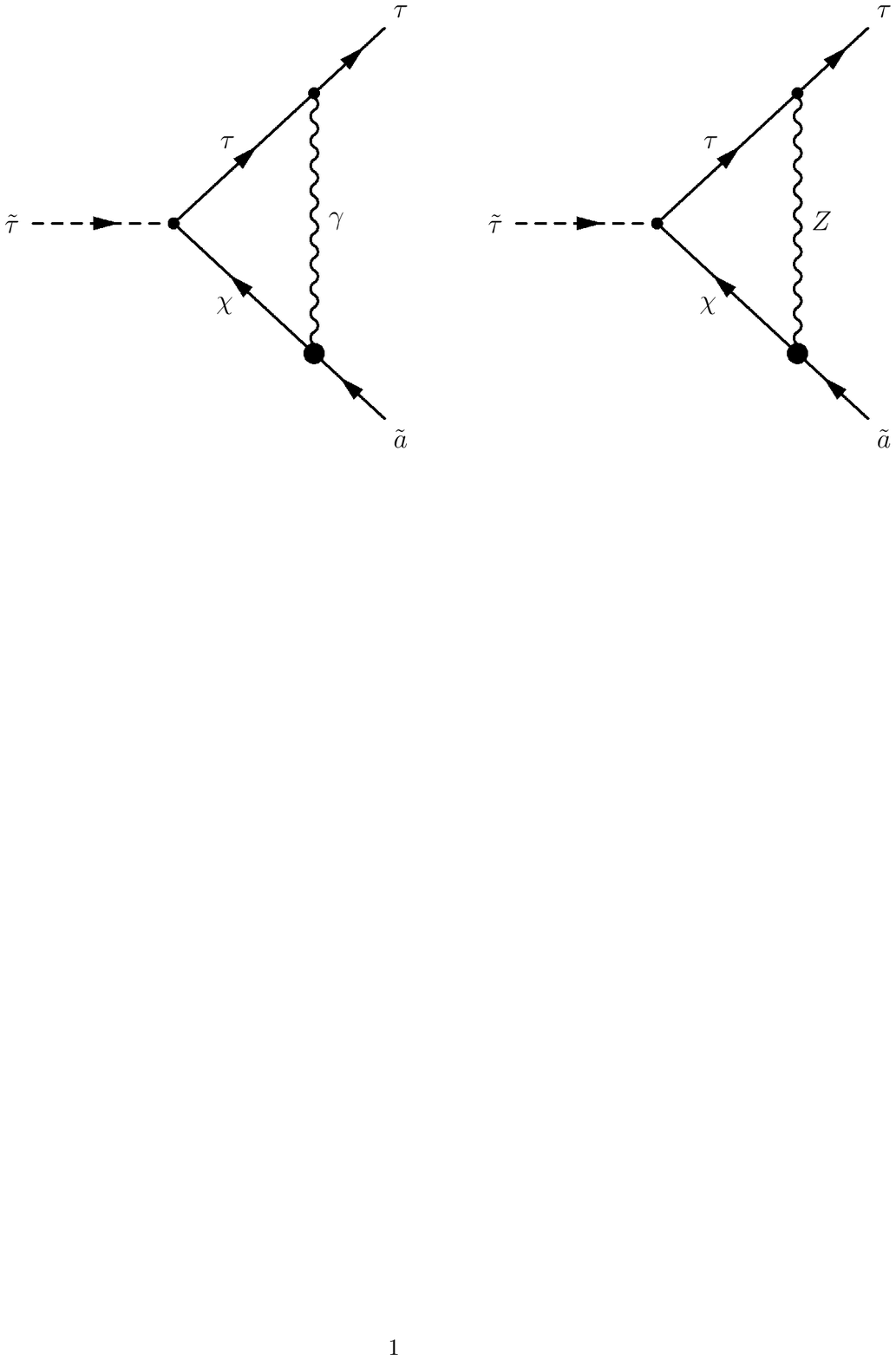, width=5in}      
    \caption{The Feynman diagrams giving the dominant contribution to
             the stau--tau--axino coupling.}
    \label{fig:loop6}
}

In the following we present the result for the loop integration. We
keep only the dominant contribution and regulate its logarithmic 
divergence with the cut--off $\fa$, in the spirit of~\cite{crs1}.

Consider first the loop containing a photon. This diagram gives rise to the 
following effective vertex involving an axino $\axino$ with momentum
$k_1$, a tau $\tau$ with momentum $k_2$ and a stau $\stau_{R/L}$ 
with momentum $k_1+k_2$,
\beq
i  \bar u_{\tau} (k_2) 
(G^\gamma_{L/R} (k_1, k_2) P_{L/R}+\widetilde G^\gamma_{R/L} (k_1,
k_2) P_{R/L})  \gamma_5 v_{\axino} (k_1) 
\eeq
The ``gauge'' couplings $G^\gamma_{L/R}$ are as follows 
\bea
G^\gamma_L &\simeq& - 
\sum_{i=1}^4 { 3\sqrt{2}\alpha_{em}^2  \over 8 \pi^2}
\({\cayy Z_{1i} Z_{i1} \over \cos^2 \theta_W}\)
\({m_{\chi_i}\over \fa}\)
\log\({\fa\over m_{\chi_i}}\) \\
G^\gamma_R &\simeq& 
+ \sum_{i=1}^4
{ 3\sqrt{2}\alpha_{em}^2  \over 16 \pi^2}
\({\cayy  Z_{1i}\over \cos^2 \theta_W}\)
\( Z_{i1} + {Z_{i2}\over \tan \theta_W}\)
\({m_{\chi_i}\over \fa}\)
\log\({\fa\over m_{\chi_i}}\),
\eea
where $i=1,\ldots,4$ denotes the four neutralino species and $Z_{ij}$ is the
matrix that changes basis to the neutralino mass eigenstates. 

The ``Yukawa'' couplings $\widetilde G_{R/L}$ arise due to the higgsino content
of the intermediate neutralino and are given by 
\beq
\widetilde G^\gamma_{R/L} \simeq - 
\sum_{i=1}^4{3 \sqrt{2} \alpha_{em}^2 \over 8 \pi^2}
\({C_{aYY} Z_{1i} } \over {\cos^2\theta_W}\)
\({Z_{i3} m_{\tau} \over 2 
m_W \tan \theta_W\cos\beta} \)
\({m_{\chi_i}\over \fa}\)
\log\({\fa\over m_{\chi_i}}\).
\eeq
The analogues of these $P_{R/L}$ couplings were not considered
in the case of the  squark--quark--axino coupling in~\cite{crs1}
because the dominant diagrams there involved an internal gluino 
instead of the neutralino.

In the calculation of $\widetilde G^\gamma_{R/L}$, $m_{\tau}$ appears in
the stau--tau--neutralino coupling but has been neglected in the
propagator and in the final state mass. This is justified because the
dominant scales in the loop and the kinematics, respectively
$m_{\chi}$ and $m_{\stau}$, are both much larger than
$m_\tau$. Moreover, since $m_{\tau}/m_W\simeq0.02$ and $Z_{1i}$ and
$Z_{i3}$ cannot both be close to $1$, we conclude that the effect of
the tau mass is small also in the coupling and that $\widetilde
G^\gamma_{R/L}$ can safely be neglected, as long as we assume
$\tan\beta < 50 $. 

We turn now to the loop containing a $Z$--boson instead of a photon. The
dominant contribution to the loop integral comes from the high energy
scale $\fa$, and is the same for both loops. In fact, adding
a $Z$--boson mass term to the denominator leads to a suppression
relative to the photon case of only a few percent, which we neglect.
Therefore, the only difference with respect to the photon loop is in
the couplings of the $Z$--boson to the tau, and to the neutralino and
the axino.  Note also that the $\widetilde G^Z_{R/L}$ couplings are
suppressed in the same way as the corresponding $\widetilde
G^\gamma_{R/L}$ and therefore will be neglected.

In summary, the two loop diagrams together lead to the following
effective $\stau_{R/L}-\tau-\axino$ couplings
\bea
\label{glsum:eq}
G_L &\simeq& - 
\sum_{i=1}^4 { 3\sqrt{2}\alpha_{em}^2  \over 8 \pi^2}
\({\cayy Z_{1i} Z_{i1} \over \cos^2 \theta_W}\) \( 1- \tan^2 \theta_W \)
\({m_{\chi_i}\over \fa}\)
\log\({\fa\over m_{\chi_i}}\) \\
& & \nonumber \\
G_R &\simeq& + \sum_{i=1}^4
{ 3\sqrt{2}\alpha_{em}^2  \over 16 \pi^2}
\({\cayy  Z_{1i}\over \cos^2 \theta_W}\) 
\({3\over 2}-{1\over 2} \tan^2 \theta_W\) \nonumber\\
& & \times
\( Z_{i1} + {Z_{i2}\over \tan \theta_W}\)
\({m_{\chi_i}\over \fa}\)
\log\({\fa\over m_{\chi_i}}\).
\label{grsum:eq}
\eea

In general, the NLSP is a linear combination of the two staus,
$\stau_1 = \cos \theta_{\stau} \,\stau_L+
\sin \theta_{\stau} \,\stau_R$.
Therefore, the photon loop leads to the following effective 
$\stau_1-\tau-\axino$ coupling
\beq
i  \bar u_{\tau} (k_1) 
\left[\sin \theta_{\stau}\, G_{L}(k_1,k_2) P_{L}+\cos
  \theta_{\stau} \,G_{R}(k_1,k_2) P_{R}\right] 
\gamma_5 v_{\axino} (k_1)
\eeq

In the following, we consider the consequences of this effective
vertex for the stau decay.

\subsection{Light Stau Decay}

Through the effective vertex described above, a
$\stau_1$ can decay into an axino and a tau, with a width
\beq
\Gamma_{\stauone\rightarrow\, \axino\,\tau}\simeq\frac{m_{\stauone}}{16 \pi} 
\left(\sin^2 \theta_{\stau}\vert\,G_L\vert^2+
\cos^2 \theta_{\stau}\vert \,G_R\vert^2\right).
\eeq

In order to get a rough estimate of the decay rate without considering
the whole neutralino sector, note that the two couplings contain the terms
$Z_{1i} m_{\chi_i} Z_{i1}$ and $Z_{1i} m_{\chi_i} Z_{i2}$.
The quantity $\Sigma_i\, Z_{1i} m_{\chi_i} Z_{i1}$
is simply $M_{11}$ of the undiagonalized neutralino mass matrix in the
flavor basis, which is just $M_1$. Similarly, 
$\Sigma_i\, Z_{1i} m_{\chi_i} Z_{i2}$ is $M_{12}$, which is zero. 
In the expressions for $G_L$
and $G_R$ above, these sums are modified slightly by the 
logarithmic factors, but as a rough estimate we can assume that
they do not modify strongly the cancellation and ignore the 
$Z_{i2}$ term in $G_R$. Then we have
\beq
G_R \sim -{G_L\over 4}\frac{3-\tan^2 \theta_W}{1-\tan^2 \theta_W}
\simeq -0.96\, G_L
\eeq
and so
\bea
\Gamma_{\stauone\rightarrow\,  \axino\, \tau}&\sim&
\frac{m_{\stauone}}{16 \pi}(G_L)^2 (1-0.069 
\cos^2 \theta_{\stau}) 
\nonumber \\
&\sim&
\frac{m_{\stauone}}{16 \pi}(G_L)^2.
\eea
This implies that the width of the lightest stau does not depend too much
on its composition. By replacing $Z_{1i} m_{\chi_i} Z_{i1}$ with $M_1$ in 
the expression for $G_L$, and noticing that the logarithm has a fairly 
weak dependence on the neutralino masses, we find that 
\beq
G_L
\sim - { 3\sqrt{2}\alpha_{em}^2  \over 2 \pi^2}
\({\cayy \over \cos^2 \theta_W}\)\left({1-\tan^2
  \theta_W}\right)\({M_1\over \fa}\) 
\log\({\fa\over m_{\chi}}\).
\eeq
This gives us the following order of magnitude estimate for the lifetime
of the lighter stau
\bea
\label{eq:staulife}
\Gamma_{\stauone\rightarrow\,  \axino\, \tau}
&\sim& 
\frac{m_{\stauone}}{16 \pi}(G_L)^2 \\
&\sim&
{ 9\alpha_{em}^4  \over (2 \pi)^5}
\({\cayy^2 \over \cos^4 \theta_W}\)({1-\tan^2 \theta_W})^2
\({M_1\over \fa}\)^2
\log^2\({\fa\over m_{\chi}}\)m_{\stauone}  \nonumber \\
&\sim&
\left(7.14\,\textrm{sec}\right)^{-1}~ 
\cayy^2 \({M_1\over 100\gev}\)^2
\({10^{11}\gev \over \fa}\)^2 \({m_{\stauone} \over 100\gev}\).
\nonumber
\eea
So the lifetime of the light stau can be larger than $1 \sec$
and in the case of NTP the decay takes place during or after 
nucleosynthesis.

Note that the stau can also decay into a tau, an axino and a photon
through the effective tree diagram in Fig.~\ref{fig:loop7}.
\FIGURE[t]{\hspace*{1in}
    \epsfig{file=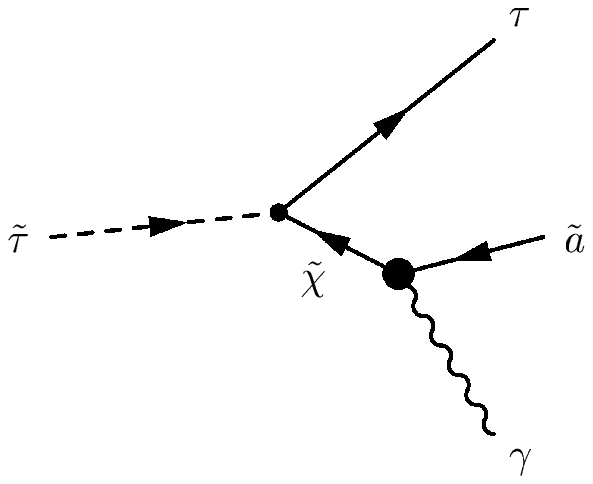, width=2.5in} \hspace*{1in}   
    \caption{Three--body decay of a stau into an axino, a tau and a photon.}
    \label{fig:loop7}
}
Even though this decay mode is not suppressed by loop factors, 
it gives a much smaller contribution to the stau width, 
\bea
\Gamma_{\stauone\rightarrow\,  \axino\, \tau\, \gamma} 
&\sim & 
{12 \alpha_{em}^3 \over (4 \pi)^4 }  
~\({C_{aYY}^2} \over {\cos^4\theta_W}\)
~\({m^3_{\stauone}\over \fa^2}\)
~F \({m_{\chi}^2} \over {m_{\stauone}^2} \) \nonumber \\
&\leq&
\left(41 \,\textrm{sec}\right)^{-1} \, C_{aYY}^2\,
\({m_{\stauone} \over 100 \gev}\)^3
\({10^{11}\gev \over \fa}\)^2\; ,
\eea
where $F(x)$ is the same as for the squark decay \cite{crs1},
\beq
F(x) =  {1\over 6} - {13 x\over 4} + {7 x^2 \over 2} +
{x\over 2} (x-1) (7 x-3) \log \[1- {1\over x}\],
\eeq
and in the last estimate we have taken the maximal value for
$F$ at $m_{\chi} = m_{\stauone} $. Note that $F(x)$ drops
fast for increasing $x$ and so we can conclude that the three 
body decay can safely be neglected.

\subsection{Constraints from Nucleosynthesis}

The stau lifetime is of the order of one second and therefore if a
large population of NLSP $\stau_1$ remains after freeze-out when NTP
is dominant, then their decays could be in conflict with predictions
of nucleosynthesis.  In fact, the produced $\tau$ immediately decays with the
lifetime of $3 \times 10^{-13}\sec$ mostly into light mesons, and
hadronic showers from this secondary decays can destroy light
elements.  Fortunately the stau lifetime is much shorter than
$10^4\sec $ and so the much more stringent bounds coming from
photo--destruction, which strongly constrained the gravitino LSP
scenario \cite{eoss04-gravitino}, are automatically avoided here.

Hadronic showers resulting from decays of non--relativistic particles
of a lifetime shorter than $10^2\sec $ have been studied
in~\cite{kohri01}. 
Although in this case it is actually the relativistic tau that decays, 
we can still use the results of the previous 
analysis if we instead consider the stau as the initial
particle. Below we will follow the procedure used before in~\cite{ckkr}.
 
Since the decay of the stau into an on--shell tau is by far the dominant 
channel, we can use the approximation
\beq
\textrm{Br}(\stau_1 \to q\, \qbar)\simeq\textrm{Br}(\tau \to q\, \qbar)
\simeq 0.63,
\eeq
where $\textrm{Br}(\tau/\stauone \to q \qbar)$ is the hadronic branching ratio 
of the tau/stau. Using this expression, we can write the bound from hadronic 
destruction of the light elements as follows 
\beq
\textrm{Br}(\tau \to q\, \qbar) Y_{\stauone} (T_F)
< (B_h Y)_{max} (\tau_{\stauone}),
\eeq
where $Y_{\stauone} = n_{\stauone}/s$ is the number density of 
$\stau_1$ divided by the entropy density and the function
$(B_h Y)_{max} (\tau_{\stauone}) $ can be read out from Fig.~12 of 
\cite{kohri01}. 
So, for example, for a stau with a mass of $100\gev $ and a lifetime
$\tau_{\stauone} = 7 \sec$, we have the bound 
$Y_{\stauone} < 0.5 \times 10^{-12} $. Note that the bound disappears
completely for lifetimes shorter than $0.04\sec $ and has a weak
dependence also on the decaying particle mass \cite{kohri01}.

To consider the most stringent scenario, we require that the out of 
equilibrium decay of the light staus is responsible for the production 
of axinos in sufficient numbers to match the present DM abundance.
Since each stau produces one axino, we have (for non--thermal production)
$Y_{\axino}^{NTP}=Y_{\stauone}(T_F)$, and therefore 
\beq
Y_{\axino}^{NTP} < 1.6\; (B_h Y)_{max} (\tau_{\stauone})
\eeq
If axinos produced are to make a sufficient contribution to the energy density
to be cosmological dark matter, they must satisfy the lower bound
\beq
\maxino Y_{\axino} > 0.34 \ev ,
\eeq
corresponding to $\Omega_{\axino} h^2 = 0.095 $ \cite{WMAP}.
In the low $T_R$ limit (where
thermally--produced axinos do not make a significant contribution), when
the $\stau_1$ is the NLSP, axinos must therefore have a mass greater than 
the following bound in order to be the dominant part of the dark matter
\beq
\maxino > 0.2 \ev  {1 \over (B_h Y)_{max} (\tau_{\stauone}) }.
\label{maxinobbn:eq} 
\eeq
If $\maxino$ is lower, axinos cannot constitute the dominant component
of dark matter, which could be then made up of other species, \eg\
axions \cite{axiondm}.  To illustrate the significance of the
constraint~(\ref{maxinobbn:eq}), let us choose $\cayy=1$ and
$\fa=10^{11}\gev$. The lower bound on $\maxino$ for this case read off
from \cite{kohri01} is shown in Fig.~\ref{fig:maxmst} for $M_1=
m_{\stauone}$ (solid line) and for the CMSSM case of degenerate
neutralino and $\stauone$ (dashed line).
\FIGURE[t]{%
    \epsfig{file=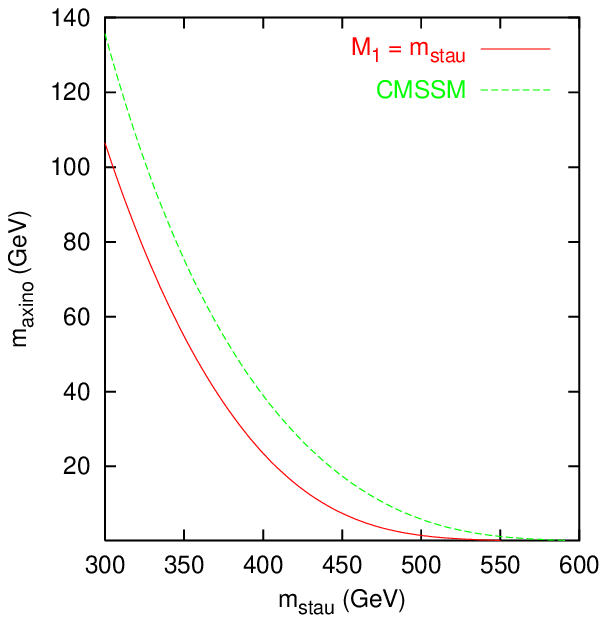, width=3.5in}      
    \caption{Lower bounds on $\maxino$ for $\cayy=1$ 
             and $\fa=10^{11}\gev$.}
    \label{fig:maxmst}
}

It is important to note that the condition~(\ref{maxinobbn:eq}) on the
axino mass does not guarantee that axinos are produced in sufficient
numbers to constitute the dark matter, but merely that, if they are
then the decays of parent staus would not disrupt
nucleosynthesis. Later we will identify regions of the parameter space
where NTP is dominant and the axino density is in the preferred
CDM range.

\section{Relic Abundance of Axinos}\label{sec:thermalaxino}

In computing the relic number density $\naxino$ of axinos we will
follow the procedure described in detail in~\cite{ckkr,crs1}.  Here we
merely briefly summarize its main points. 

As regards TP, the number density of axinos $\naxino$ can be obtained
by integrating the Boltzmann equation with both scatterings and decays
of particles in the plasma included. Since $\naxino$ is well below the
equilibrium one for $\treh \ll \fa$, we can neglect inverse
processes. In terms of yield
\begin{equation}
\ythaxino= \frac{\naxino^{\rm TP}}{s} = \sum_{i}\yadec + \sum_{i,j}\yascat,
\label{ythaxino:eq}
\end{equation}
where $s= (2\pi^2/45)\gsstar T^3$ is the entropy density, 
and normally $\gsstar=\gstar$ in the early Universe. 
By changing variables from the cosmic time $t$ to the temperature 
$T$,
we can write the two solutions of the Boltzmann differential equation 
easily in integral form
\bea
\label{eq:Ydec}
\yadec (T_{0}) &=& 
\int_{T_{0}}^{\treh}dT\,
\frac{\langle\Gamma(i\rightarrow\, \axino +\cdots)\rangle n_i}{sHT}\\
\label{eq:Yscat1}
\yascat (T_{0}) &=&
\int_{T_{0}}^{\treh}dT\,
\frac{\langle\sigma(i+j\rightarrow\, \axino +\cdots)\rangle n_in_j}{sHT}.
\eea
where we have considered the evolution from the reheating temperature 
after inflation, $\treh$, down to the present temperature $T_{0}$.

In computing $\yadec$, in addition to the previously considered decays
of gluinos~\cite{ckkr} and (s)quarks~\cite{crs1}, we now include for
consistency the decays of the staus that are generated by the
couplings~(\ref{glsum:eq}) and~(\ref{grsum:eq}), 
even though their effect is usually negligible. 
Also scatterings arising from these effective couplings
invariably give subdominant contributions~\cite{ckkr,crs1}
and will be neglected here. The relic abundance due to thermal
production is then calculated by using the formula
\beq
\maxino \yaxino \simeq 0.36 \ev \(\Omega_{\axino}^{\rm TP}h^2 \over 0.1\).
\label{eq:m-yaxino}
\eeq

In order to evaluate the axino relic abundance generated through NTP
processes we first compute the relic number density of NLSPs after
they freeze out from the plasma. In the case of the neutralino, we
include all tree--level two--body neutralino processes of
pair--annihilation and co--annihilation with the charginos,
next--to--lightest neutralinos and sleptons.  This is a standard
method which allows us to accurate compute $\abundchi$ in the usual
case when the lightest neutralino is the LSP.  We further extend the
above procedure to the case when it is the lightest stau $\stauone$
that is the NLSP. We include all stau--stau annihilation and
stau--neutralino co-annihilation processes.  In both cases we solve the
Boltzmann equation numerically and properly take into account
resonance and new final--state threshold effects.  The procedure has
been described in detail in~\cite{cmssm-rrn}.

Since all the NLSPs subsequently decay into axinos, a simple relation
holds
\beq
\Omega^{\rm NTP}_{\axino}=\frac{\maxino}{m_{\rm NLSP}}\Omega_{\rm NLSP}.
\label{omegaantp:eq}
\eeq
In the following, we will first illustrate the impact of the
cosmological constraints on the plane ($\mhalf,\mzero$) in the CMSSM
without the axino, in which case the LSP is normally the lightest
neutralino or the stau.

\section{DM in the CMSSM without the Axino LSP}\label{sect:cmssm}

In contrast to the general MSSM, mass spectra of the CMSSM are tightly
inter--related. This is because the model is defined in terms of
only the usual five free parameters: $\tanb$, the common gaugino mass
$\mhalf$, the common scalar mass $\mzero$, the common trilinear soft
scalar coupling $\azero$ and $\sgn(\mu)$ -- the sign of the
supersymmetric Higgs/higgsino mass parameter $\mu$. For a fixed value
of $\tanb$, physical masses and couplings are obtained by running
various mass parameters, along with the gauge and Yukawa couplings,
from their common values at $\mgut$ down to $\mz$ by using the
renormalization group (RG) equations.  The large top quark Yukawa
coupling tends to push $m_{H_2}^2$ to negative values around the
electroweak scale. Therefore electroweak symmetry is spontaneously
broken and $\mu^2$ is determined, but not its sign.
%
%

Experimental and cosmological constraints on the CMSSM are usually
presented in the ($\mhalf,\mzero$) plane for various
representative choices of $\tanb$ and other relevant parameters.
In this paper, we will not attempt a detailed study of the
CMSSM. Instead, we will aim at demonstrating that very
different patterns arise by assuming the standard scenario and that with
the axino LSP. 

\FIGURE[t]{
\vspace*{-0.27in}
\hspace*{-.70in}
\epsfig{file=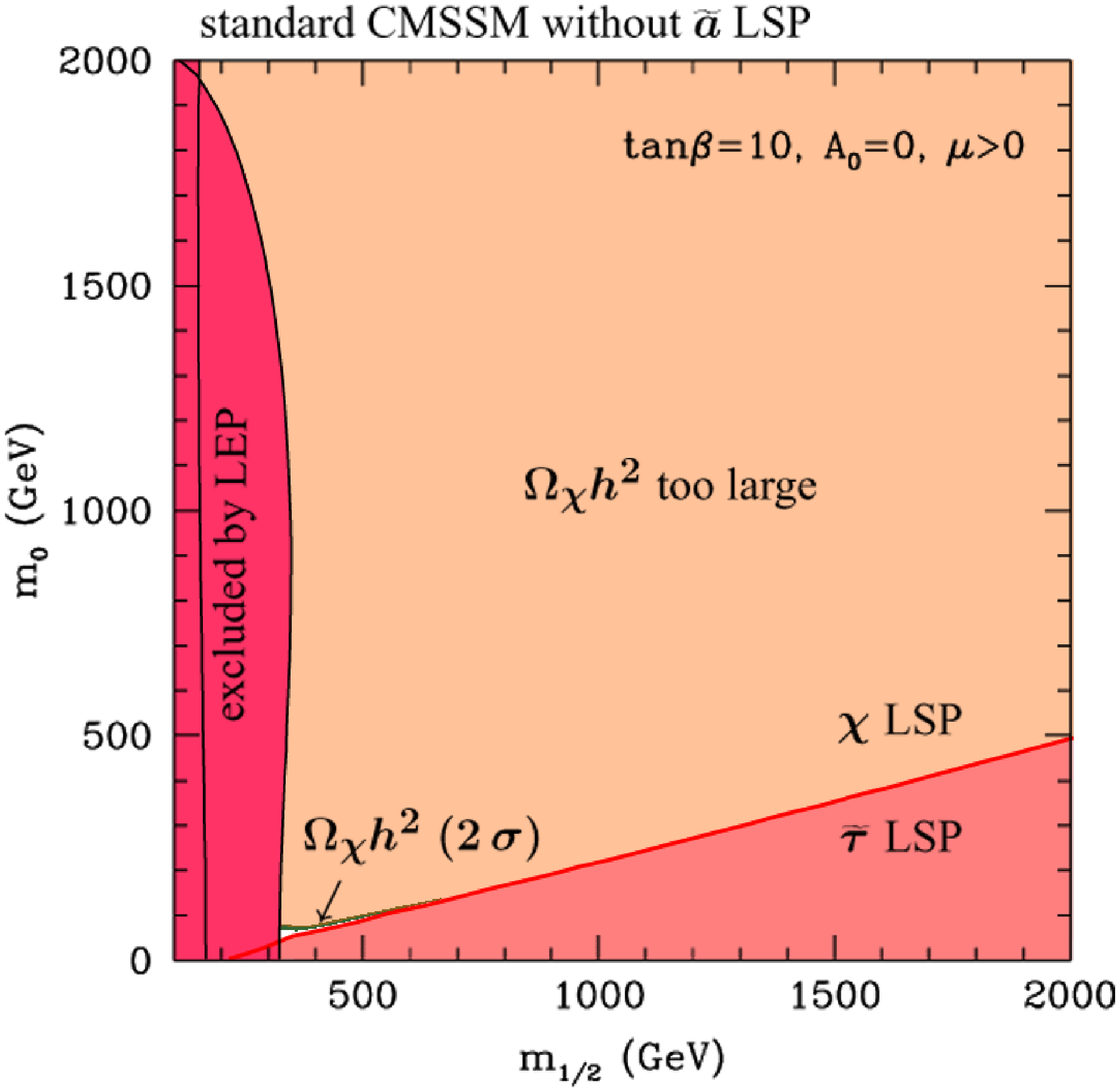,width=4.0in}
\caption{\small The ($\protect{\mhalf},\protect{\mzero}$) plane for 
        the standard scenario without the axino LSP, and for 
        $\protect{\tanb=10, \azero=0}$ and $\mu>0$. In the leftmost 
        (red) region marked ``excluded by LEP'' the chargino (to the 
        left of a solid line) or the light Higgs is too light. 
        In the tiny dark green band
        marked ``$\abundchi\;(2\,\sigma)$'' $0.095<\abundchi<0.130$,
        while in the orange region labeled ``$\abundchi$ too large''
        $\abundchi$ exceeds the upper limit. In the white region
        $\abundchi$ is less than the favored $2\,\sigma$ range but
        otherwise allowed.}
\label{maxino_cmssm:fig}
}

In Fig.~\ref{maxino_cmssm:fig}, we show the plane ($\mhalf,\mzero$)
for $\tanb=10$, $\azero=0$ and for $\mu>0$. In the absence of the
axino,  the lightest neutralino
is the LSP above a solid red line, while in the wedge below it the LSP is
the lighter stau $\stauone$. The red 
region on the left is excluded by LEP lower bounds on the mass of the
chargino $\mcharone>104\gev$ (far-most solid line) and the light Higgs
$\mhl\gsim114.4\gev$ \cite{LEP}.  In the CMSSM without the axino LSP, the wedge
is considered to be excluded as it would give an electrically charged
stable relic. Likewise, most of the region of neutralino LSP is
excluded due to the relic abundance $\abundchi$ exceeding current
bounds. For a stable relic we will require that its relic abundance
falls into the $2\,\sigma$ range
\begin{equation}
0.095<\abundcdm<0.130
\label{eq:wmaprange}
\end{equation}
which follows from combining WMAP results \cite{WMAP} with other recent
measurements of the CMB. In Fig.~\ref{maxino_cmssm:fig}, the
constraint~(\ref{eq:wmaprange}) is fulfilled only in a very narrow
(green) band just above and along the red line. In this case,
neutralino pair annihilation is inefficient because the sleptons and
squarks are already too heavy to reduce $\abundchi$ through
$t$--channel processes. Nor are the heavy Higgs scalars $\hh$ and
$\ha$ light enough to allow for resonant enhancements, although this
becomes the case at very large $\tanb\gsim50$ which we do not consider
here. The mechanism that comes to the rescue is the neutralino
co--annihilation with $\stauone$~\cite{efo98}, which however, for $\tanb=10$
considered here, is efficient only for $\mhalf\lsim500\gev$. In the
narrow white region below the green band, $\abundchi$ is less than the
favored range of~(\ref{eq:wmaprange}).  Except for these two, the rest
of the ($\mhalf,\mzero$) plane is cosmologically excluded.

Clearly, the standard paradigm is highly constrained. We will now
proceed to compare it with the case of axino as the LSP and CDM.  As
we will see, the form of the cosmological constraint will in general
become not only very different but also will strongly depend on the
axino mass and the reheat temperature.

\FIGURE[t!]{
\vspace*{-0.27in}
\hspace*{-.70in}
\hspace*{-0.15in}
\begin{minipage}{6.5in}
\epsfig{file=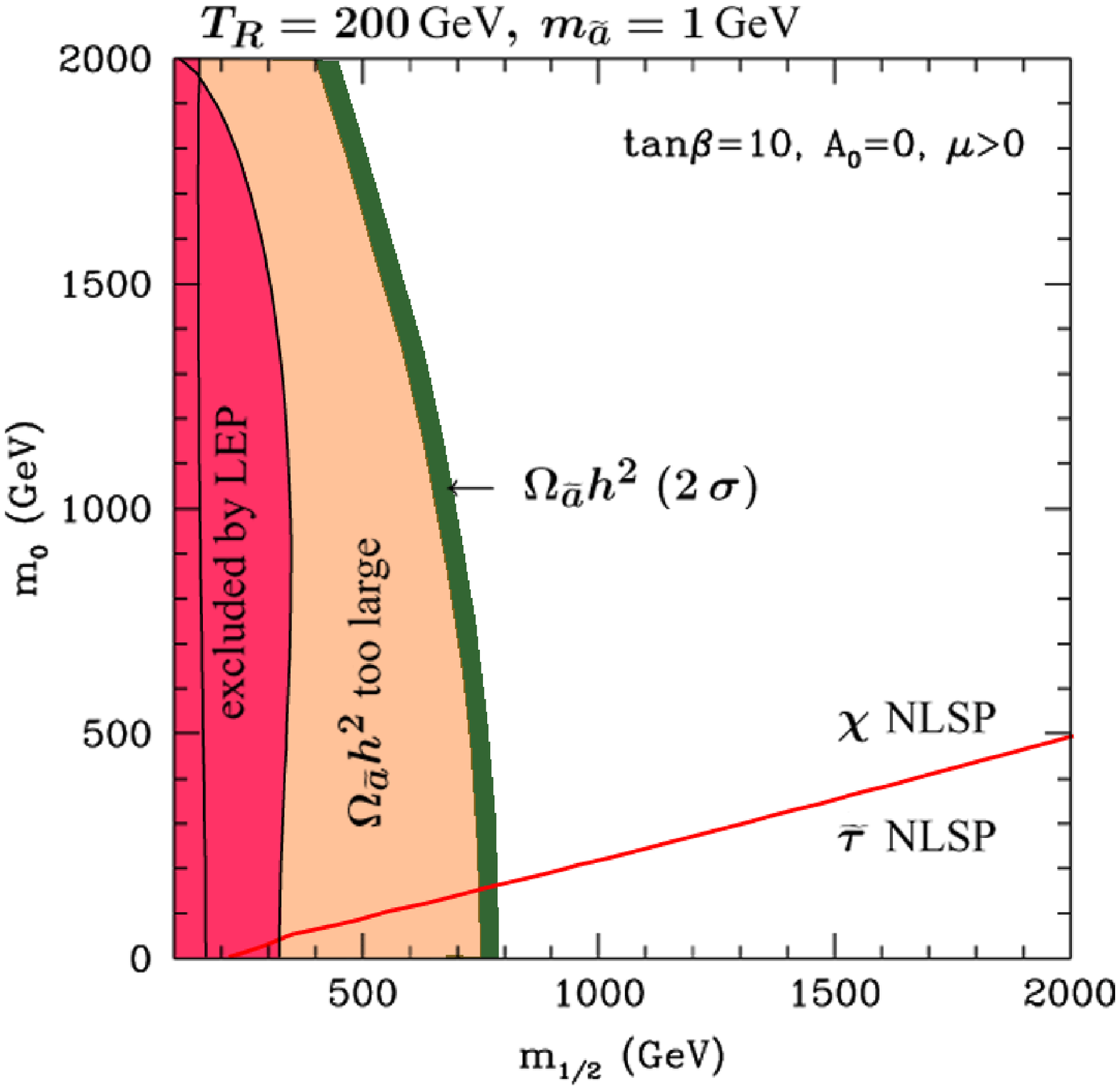,height=3.425in}
\hspace*{-0.19in}
\epsfig{file=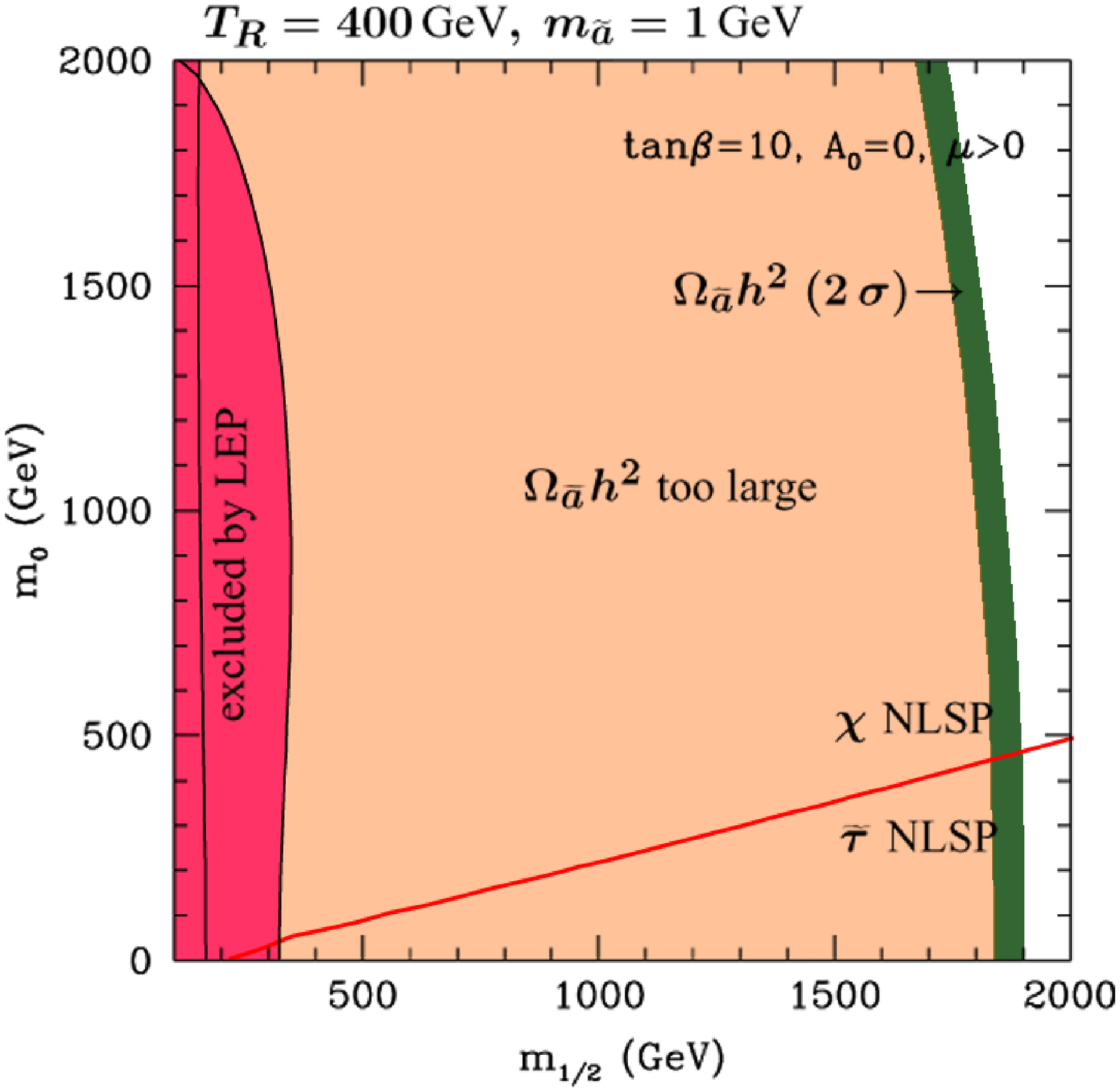,height=3.425in}
\end{minipage}
\caption{\small The ($\protect{\mhalf},\protect{\mzero}$) plane for 
        axino LSP with $\maxino=1\gev$ and for $\fa=10^{11}\gev$,
        $\protect{\tanb=10,\azero=0}$ and $\mu>0$.  In the left panel,
        $T_R=200\gev$, in the right, $400\gev$. The dark
        green (orange, white) regions correspond to
        $0.095<\abunda<0.130$ ($\abunda$ too large and excluded, too
        small but otherwise allowed).}
\label{maxino_1:fig}
}

\section{DM in the CMSSM with the Axino LSP}\label{sect:cmssmaxino}

As summarized in the Introduction, if the axino is light,
$\maxino\lsim10\mev$ (while still remaining a cold DM candidate, \ie\
$\maxino\gsim100\kev$), then thermal production is the dominant source
of axinos, and non--thermal production can be neglected.  The reheat
temperatures favored by cosmology (such as to give $\abunda\sim0.1$) are in
the few hundred $\tev$ range, independently of $\mhalf$
and $\mzero$ for any sensible values of these parameters.

If the axino mass is increased to $1\gev$, non--thermal production is
typically 
still negligible. However, the cosmologically favored range $\abunda$
then requires $\treh\lsim {\cal O} (1\tev)$ at which point squark and
slepton decays in TP start playing a major role. This can be seen in
Figs.~7 and~9 of~\cite{crs1} in the case of squark decays. In
particular, one can see in Fig.~9 of~\cite{crs1} that, at fixed
$\maxino$, as squark masses decrease, a maximum $\treh$ allowed by
$\abunda<0.13$ also decreases.  A convenient way to present this is to
invert the reasoning and to consider the cosmologically favored ranges
of squark and slepton masses at fixed $\treh$. Since these increase
with $\mhalf$ and $\mzero$, this will be reflected in cosmologically
favored regions of the ($\mhalf,\mzero$) plane at given fixed values
of $\maxino$ and $\treh$.

In the left panel of Fig.~\ref{maxino_1:fig} the ($\mhalf,\mzero$)
plane is shown for a reheat temperature of $200\gev$, while on the
right this is increased to $400\gev$. In both panels
$\maxino=1\gev$. Other relevant parameters have been set to the
following values: $\fa=10^{11}\gev$, $\tan \beta=10$, $A_0=0$, and
$\mu>0$.  The red line divides the neutralino and stau NLSP
regions. Below the line, the stau is the NLSP, while above it the
neutralino.  As in Fig.~\ref{maxino_cmssm:fig}, the red band along the
$\mzero$ axis is excluded by imposing the LEP bounds on the chargino
and Higgs mass.
The cosmologically favored range is coloured green, while the region
excluded by the requirement that axinos should not overclose the
Universe ($\abunda \lsim0.130$) is marked in orange. The white regions
are cosmologically allowed, but not favored, \ie\ $\abunda
\lsim0.095$.

In both panels, it is clear that thermally produced axinos exclude a
region closer to the $\mzero$ axis, while there is a cosmologically
favored strip further out. As $\treh$ is increased, the excluded region
grows and the favored region is pushed farther out to the right.  This
can be understood by again examining Fig.~7 of~\cite{crs1}: at fixed
sfermion and gluino masses, as $\treh$ is taken to larger values, the
corresponding yield and therefore (at fixed $\maxino$) $\abunda$
increase.  The important point to note here is that a portion of the
cosmologically favored region lies in the stau NLSP wedge, which is
traditionally thought to be excluded. 
Note that the bounds of Fig.~(\ref{fig:maxmst}) are not applicable here 
because NTP is negligible in this region and also that other BBN bounds
are irrelevant for cold axinos~\cite{ckkr}, similarly as for
neutralino CDM.

We now consider the other limit in which the axino is as heavy as
possible, but slightly lighter than the NLSP in order to avoid 
a strong phase space suppression in the NLSP decay. In this
regime non--thermal production is no longer negligible. In fact, at
low enough $\treh$, as the yield due to TP becomes too small, it is
NTP that starts providing enough axinos.  Again, we refer the reader
to Figs.~7--9 of~\cite{crs1}. We plot the case of $\maxino\sim
m_{\rm NLSP}$, in Fig.~\ref{maxino_max:fig}. In the left panel the
reheat temperature is $50\gev$, while on the right this is increased
to $200\gev$.  The other parameters are set to the same values as in
Fig.~\ref{maxino_1:fig}.

\FIGURE[t!]{
\vspace*{-0.27in}
\hspace*{-.70in}
\begin{minipage}{6.5in}
\hspace*{-0.15in}
\epsfig{file=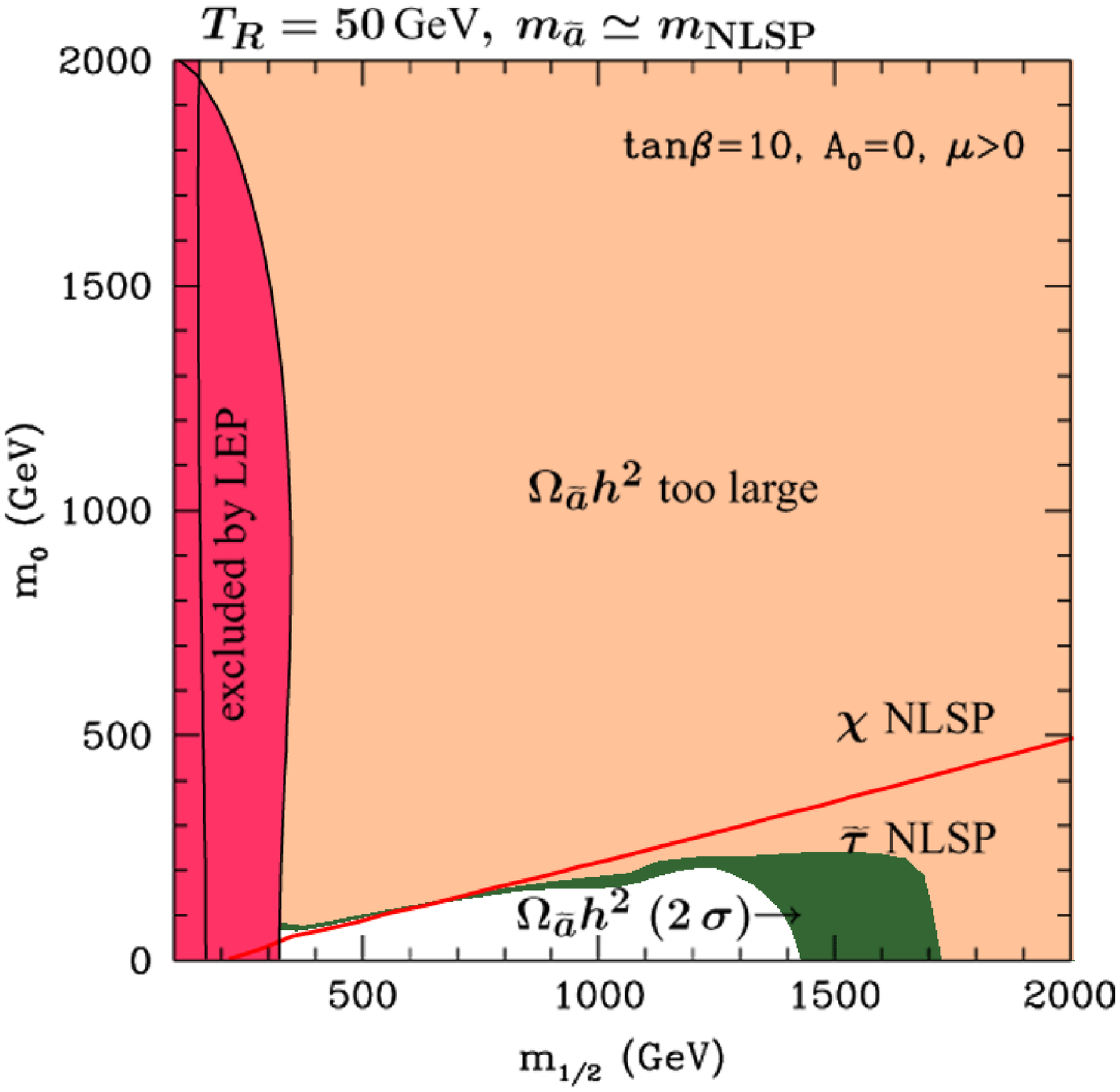,height=3.425in}
\hspace*{-0.19in}
\epsfig{file=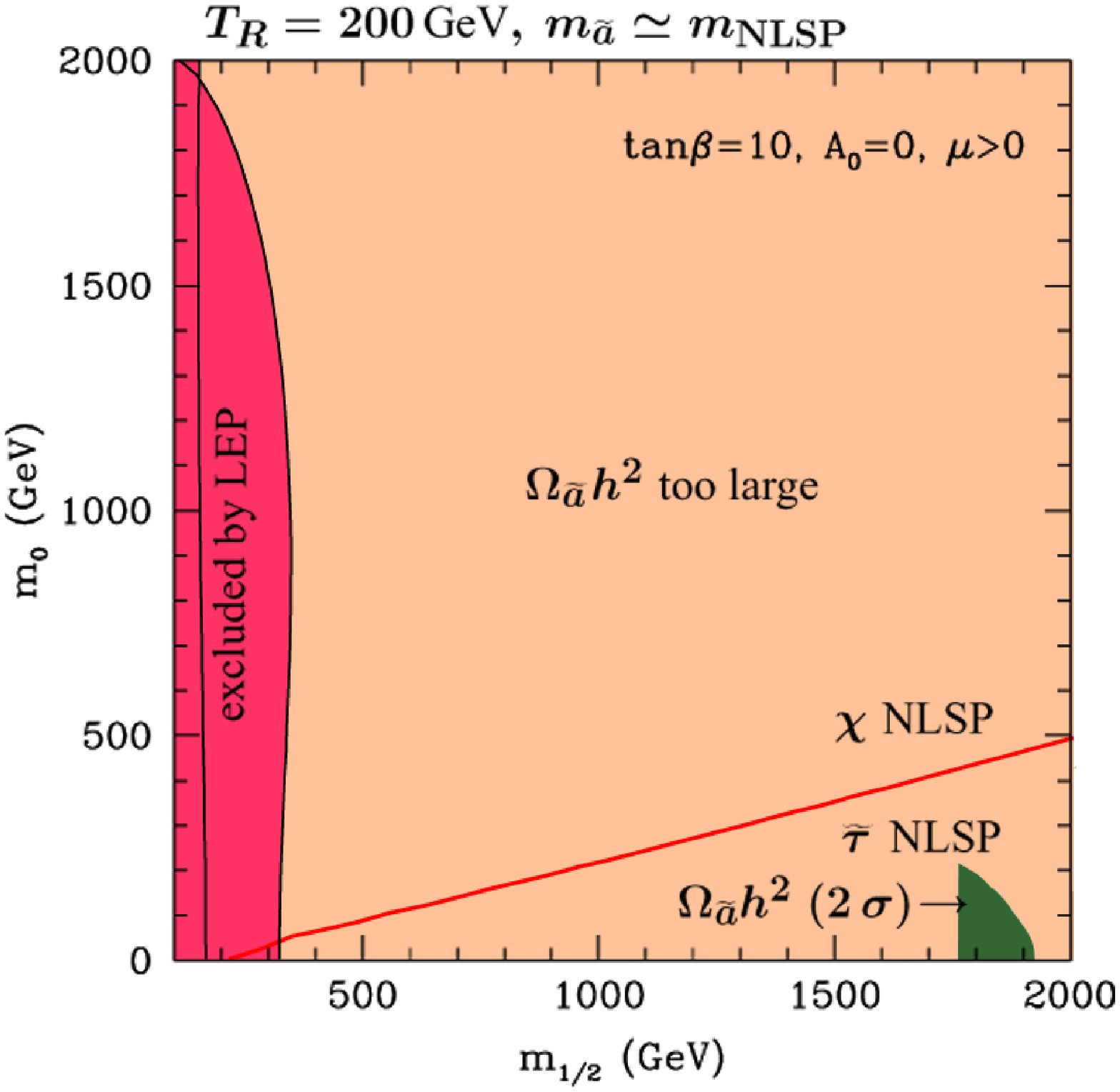,height=3.425in}
\end{minipage}
\caption{\small The ($\protect{\mhalf},\protect{\mzero}$) plane for 
        an axino only slightly less massive than the NLSP, with 
        $\fa=10^{11}\gev$,
        $\protect{\tan \beta=10}$, $A_0=0$, and $\mu>0$.  In the left
        panel, $T_R=50\gev$, in the right, $200\gev$. The notation
        follows that of Fig.~\protect{\ref{maxino_1:fig}}.}
\label{maxino_max:fig}
}

In the left panel, in which $\treh=50\gev$, the thermal
production of axinos now plays a subdominant role. This is
reflected in the fact that the green region of favored $\abunda$ in
the neutralino NLSP section, at $\mzero\sim100\gev$, coincides with
the green region of the neutralino LSP in Fig.~\ref{maxino_cmssm:fig}
as a result of the relation~(\ref{omegaantp:eq}). For the same reason, at
$\mhalf\simeq700\gev$, the green region starts to extend further down
into the stau NLSP wedge, along the line dividing the stau and
neutralino NLSP sections, as co--annihilations between the neutralino
and the stau keep the LSP density within the expected range. Further
to the right, the co--annihilation ceases to be effective but stau
decays provide enough $\abunda$ for a limited range of $\mstauone$. It is
interesting to note that the majority of the cosmologically favored
region now lies in the stau NLSP wedge. In this region the lower
bound on the axino mass given in Fig.~\ref{fig:maxmst} as CMSSM
applies and it is in our case satisfied.
Note also, that even at such low temperature, the NLSP is completely 
thermalized in the whole ($\protect{\mhalf},\protect{\mzero}$) plane.

In the right panel, in which $\treh =200\gev$, TP can no longer be
neglected, and in fact thermally produced axinos overclose the
Universe if the neutralino is the NLSP. However, there is still a
green region at $\mhalf\simeq 1900\gev$ in the stau NLSP region, in
which TP and NTP axinos together form the cosmologically favored
abundance of CDM.

The left panel of Fig.~\ref{maxino_1:fig} and the right panel of
Fig.~\ref{maxino_max:fig} allow one to see the effect of increasing
$\maxino$ at fixed $\treh$. The cosmologically favored band due to TP
axinos moves to the right along the $\mhalf$ axis, in a similar way to
the effect of increasing $\treh$. This amounts to increasing sfermion
and gluino masses which in turn will increase the importance of NTP
relative to TP, as can be seen in Fig.~7 of~\cite{crs1}. 
As a result, in the right
panel of Fig.~\ref{maxino_max:fig}, NT production gives too much relic
density in the cosmologically favored band for TP axinos, except for
the small region in the corner of small $\mzero$ and large
$\mhalf$. In the left part of this region, TP axinos provide the
required abundance, and as you move 
to the right within the region TP axinos become subdominant and NTP axinos
provide the necessary abundance, before eventually producing too many
of them.

If one increased the reheat temperature further up one would exclude the
region where NTP axinos can be CDM, and pushed the region where TP
axinos can perform this function to values of $\mhalf$ larger than
those plotted here.  On the other hand, decreasing the reheat
temperature below a few~$\gev$ would have the effect of suppressing
the relic abundance of NLSPs, so that neither TP nor NTP could provide
axino CDM. We will not investigate this possibility here.

\section{Conclusion}

The Constrained MSSM provides a popular and predictive framework for
analyzing properties of low--energy SUSY. While theoretical
assumptions and experimental bounds from LEP and $b\to
s\gamma$ provide significant constraints on the parameter space, it is
the cosmological relic density of the LSP as cold DM and the
requirement of its electric neutrality that rule out most of the
($\mhalf,\mzero$) plane and, for $\tanb\lsim50$, allow for only a tiny
region in the plane. 

In this paper, we have considered the cosmological bounds on the
scenario in which the LSP is not the usual neutralino but instead the
axino. Our purpose was not to provide an exhaustive scan of the CMSSM but to
illustrate with a few examples the very different patterns that
arise.  We concentrated on two cases of a fairly light axino
$\maxino=1\gev$ and of $\maxino\sim m_{\rm NLSP}$ where the NLSP was
either the neutralino or the lighter stau. In computing the axino
relic abundance we included both thermal and non--thermal production
processes.  Our results are a function of $\treh$ which in all cases
had to be rather low, below a few hundred $\gev$ in order to give the
right amount of axino CDM. We further included constraints from
nucleosynthesis which are important for lighter axinos and staus.

The cosmologically favored regions consistent with axino CDM relic
abundance are in general very different from the usual
scenario. Depending on $\maxino$ and/or $\treh$, basically nearly any
point in the ($\mhalf,\mzero$) plane can become cosmologically
allowed. This is true for both the regions where the NLSP is the
neutralino or the lighter stau.  This significantly relaxes the
cosmological bounds obtained in the absence of the axino.

From the point of view of collider phenomenology, with the lifetime of
${\cal O}(1\sec)$, the NLSP would appear stable in a
detector. However, if the NLSP is the neutralino, then very likely one
would find its relic abundance (calculated under the assumption
that it is the true LSP) to significantly exceed $0.13$. The NLSP could also be
electrically charged (stau), or even colored (stop) -- the possibility
we have not investigated here. 
In conclusion, in collider studies of the CMSSM parameter space, both
the apparently cosmologically excluded bulk of the $(\mhalf,\mzero)$
plane where the neutralino is the NLSP and the wedge at its bottom where
the lighter stau is the NLSP, should be given equal
attention as the narrow regions preferred by the standard paradigm.

\acknowledgments
LC would like to thank T. Asaka, W. Buchm\"uller and G. Moortgat--Pick
for useful discussions and the Physics Department of Lancaster University 
for their kind hospitality. \linebreak
During the course of this work RRdA was 
funded by the "EU Fifth Framework Network 
'Supersymmetry and the Early Universe' (HPRN-CT-2000-00152)" and
MS was supported by a PPARC studentship.


\begin{thebibliography}{99}
 
\bibitem{MSSM-dm} G.~Jungman, M.~Kamionkowski and K.~Griest,
\prep{267}{1996}{195}.  

\bibitem{CMSSM-dm} For recent reviews, see, \eg, 
L.~Roszkowski, 
{\it Pramana} {\bf 62} (2004) 389 [arXiv:hep-ph/0404052];
C.~Mu{\~n}oz, 
to appear in Int.~J.~Mod.~Phys.~A., arXiv:hep-ph/0309346;
A.~B.~Lahanas, N.~E.~Mavromatos, D.~V.~Nanopoulos, 
\ijmpd{12}{2003}{1529} [arXiv:hep-ph/0308251].

\bibitem{na92}
P.~Nath and R.~Arnowitt, 
\prl{70}{1993}{3696} [arXiv:hep-ph/9302318].

\bibitem{rr93} 
R.~G.~Roberts and L.~Roszkowski, 
\plb{309}{1993}{329} [arXiv:hep-ph/9301267].

\bibitem{kkrw94}
G.~L.~Kane, C.~Kolda, L.~Roszkowski, and J.~D.~Wells, 
\prd{49}{1994}{6173} [arXiv:hep-ph/9312272]. 

\bibitem{efo98} 
J.~R.~Ellis, T.~Falk and K.~A.~Olive,
\plb{444}{1998}{367} [arXiv:hep-ph/9810360].

\bibitem{eos03}
C.~Boehm, A.~Djouadi and M.~Drees,
\prd{62}{2000}{035012} [arXiv:hep-ph/9911496]; 
J.~R.~Ellis, K.~A.~Olive and Y.~Santoso, 
\app{18}{2003}{395} [arXiv:hep-ph/0112113].

\bibitem{ch-lsp}
For electrically charged relics, see \eg\
T.~K.~Hemmick et al., \prd{41}{1990}{2074};
P. Verkerk et al., \prl{68}{1992}{1116};
P.~F.~Smith, {\it Contemp. Phys.} {\bf 29} (1998) 159.
%
For strongly interacting relics, see \eg\
D.~Javorsek~II et al., \prd{64}{2001}{012005} and
\prl{87}{2001}{231804}. 

\bibitem{pq}
R.~D.~Peccei and H.~R.~Quinn, \prl{38}{1977}{1440} and \prd{16}{1977}{1791}.

\bibitem{kmn}
J.~E.~Kim, A.~Masiero and D.~V.~Nanopoulos, \plb{139}{1984}{346}.    
 
\bibitem{bgm}
S.~A.~Bonometto, F.~Gabbiani and A.~Masiero,
\plb{222}{1989}{433} and \prd{49}{1994}{3918}
[arXiv:hep-ph/9305237].

\bibitem{rtw}
K.~Rajagopal, M.~S.~Turner and F.~Wilczek, 
\npb{358}{1991}{447}.

\bibitem{ckkr}
L.~Covi, H.~B.~Kim, J.~E.~Kim and L.~Roszkowski, 
\jhep{05}{2001}{033} [arXiv:hep-ph/0101009].

\bibitem{ay00}
T.~Asaka and T.~Yanagida, 
\plb{494}{2000}{297} [arXiv:hep-ph/0006211].

\bibitem{ckr}
L.~Covi, J.~E.~Kim and L.~Roszkowski, 
\prl{82}{1999}{4180} [arXiv:hep-ph/9905212].

\bibitem{crs1} 
L.~Covi, L.~Roszkowski and M.~Small, 
\jhep{07}{2002}{023} [arXiv:hep-ph/0206119].

\bibitem{ckl00}
E.~J.~Chun, H.~B.~Kim and D.~H.~Lyth, 
\prd{62}{2000}{125001} [arXiv:hep-ph/0008139].

\bibitem{kk02}
H.~B.~Kim and J.~E.~Kim, 
\plb{527}{2002}{18} [arXiv:hep-ph/0108101];
D.~Hooper and  L.-T.~Wang,
arXiv:hep-ph/0402220.

\bibitem{axinomass}
E.~J.~Chun, J.~E.~Kim and H.~P.~Nilles, 
\plb{287}{1992}{123} [arXiv:hep-ph/9205229];
E.~J.~Chun and A.~Lukas, 
\plb{357}{1995}{43} [arXiv:hep-ph/9503233];
P.~Moxhay and K.~Yamamoto, \plb{151}{1985}{363};
T.~Goto and M.~Yamaguchi, \plb{276}{1992}{103}.

\bibitem{ksvz}
J.~E.~Kim, \prl{43}{1979}{103};
M.~A.~Shifman, V.~I.~Vainstein and V.~I.~Zakharov, 
\npb{166}{1980}{4933}.


\bibitem{lep-ch-lsp}
LEPSUSYWG, ALEPH, DELPHI, L3 and OPAL experiments, 
note LEPSUSYWG/02-05.1 
(http://lepsusy.web.cern.ch/lepsusy/Welcome.html). 

\bibitem{eoss04-gravitino}
J.L.~Feng, A.~Rajaraman and F.~Takayama,
\prl{91}{2003}{011302} 
[arXiv:hep-ph/0302215] and 
\prd{68}{2003}{063504} 
[arXiv:hep-ph/0306024];
J.~Ellis, K.~A.~Olive, Y.~Santoso and V.~Spanos, 
arXiv:hep-ph/0312262.

\bibitem{thermal-grav}
M.~Bolz, A.~Brandenburg and W.~Buchm\"uller, 
\npb{606}{2001}{518} [arXiv:hep-ph/0012052].

\bibitem{axion}
S.~Weinberg, \prl{40}{1978}{223};
F.~Wilczek, \prl{40}{1978}{279}.

\bibitem{axionreviews:cite}
J.~E.~Kim, \prep{150}{1987}{1};
M.S.~Turner, \prep{197}{1990}{67};
G.G.~Raffelt,  \prep{198}{1990}{1};
P.~Sikivie, \npps{87}{2000)}{41}
[arXiv:hep-ph/0002154].

\bibitem{qcdanomaly:cite}
W.~A.~Bardeen, S.-H.~H.~Tye, \plb{74}{1978}{229};
V.~Baluni, \prd{19}{1979}{2227}.

\bibitem{dfsz}
M.~Dine, W.~Fischler and M.~Srednicki, 
\plb{104}{1981}{99};
A.~P.~Zhitnitskii, \sjnp{31}{1980}{260}.

\bibitem{susyaxion}
H.~P.~Nilles and S.~Raby, \npb{198}{1982}{102};      
J.~E.~Kim and H.~P.~Nilles, \plb{138}{1984}{150}.
 
\bibitem{anom-ren} 
S.~L.~Adler and W.~A.~Bardeen, \prd{49}{1994}{551}.

\bibitem{coupling}
J.~E.~Kim, 
\prd{58}{1998}{055006} [arXiv:hep-ph/9802220].
See, also,
D.~B.~Kaplan, \npb{260}{1985}{215};
M.~Srednicki, \npb{260}{1985}{689}.

\bibitem{martinaxino}
S.~P.~Martin, \prd{62}{2000}{095008}
[arXiv:hep-ph/0005116].

\bibitem{kohri01}
K.~Kohri, 
\prd{64}{2001}{043515} [arXiv:astro-ph/0103411]. 

\bibitem{WMAP}
D.~N.~Spergel, {\it et~al.}, 
\apj{148}{2003}{175} [arXiv:astro-ph/0302209].

\bibitem{axiondm}
J.~Preskill, M.~B.~Wise and F.~Wilczek, 
\plb{120}{1983}{127};
L.~F.~Abbott and P.~Sikivie, \plb{120}{1983}{133};
M.~Dine and W.~Fischler, \plb{120}{1983}{137}.

\bibitem{cmssm-rrn} 
L.~Roszkowski, R.~Ruiz de Austri and T.~Nihei,
\jhep{08}{2001}{024} [arXiv:hep-ph/0106334]; 
T.~Nihei, L.~Roszkowski, R.~Ruiz de Austri, 
\jhep{05}{2001}{063} [arXiv:hep-ph/0102308];
\jhep{03}{2002}{031} [arXiv:hep-ph/0202009] and 
\jhep{07}{2002}{024} [arXiv:hep-ph/0206266].

\bibitem{LEP}
LEPSUSYWG, ALEPH, DELPHI, L3 and OPAL experiments, note LEPSUSYWG/01-03.1
(http://lepsusy.web.cern.ch/lepsusy/Welcome.html). 

\end{thebibliography}
\end{document}